\documentclass[10pt,conference]{IEEEtran}

\usepackage{cite}
\usepackage{graphicx}
\usepackage[table]{xcolor}
\usepackage{amsmath,amssymb}
\usepackage{url}
\usepackage{booktabs}
\usepackage{array}
\usepackage{tikz}
\usetikzlibrary{arrows.meta,positioning}

\newcommand{\tool}{\textsc{Schwarz}}

\title{\tool: Solver-Aware Agentic Program Verification}
\author{
\IEEEauthorblockN{Jingyu Ke}
\IEEEauthorblockA{Shanghai Jiao Tong University\\
windocotber@sjtu.edu.cn}
\and
\IEEEauthorblockN{Ling-I Wu}
\IEEEauthorblockA{Shanghai Jiao Tong University\\
edithwuly@sjtu.edu.cn}
\and
\IEEEauthorblockN{Guoqiang Li}
\IEEEauthorblockA{Shanghai Jiao Tong University\\
li.g@sjtu.edu.cn}
}

\begin{document}

\maketitle

\begin{abstract}
Agentic verification systems can often generate source-level specifications
that look plausible, but plausibility is not enough: the verifier must still
turn those specifications into SMT obligations that the solver can prove. When
this step fails, current LLM-driven loops usually expose only a coarse verifier
error, timeout, or unknown solver result. The model cannot tell whether the
specification is wrong, a helper lemma is missing, the proof context contains
irrelevant facts, or the obligation needs a different theory view. This paper
presents \tool{}, an agentic verification harness that makes SMT-backed proof
failure local, checkable, and repairable. \tool{} turns failed verification into
obligation-local repair tasks: program-point snapshots expose checked facts at a
boundary, local lemmas let the agent propose missing proof steps, and
theory-aware solver policies guide the agent toward
solver-friendly formulations for numeric, quantified, memory, and floating-point
obligations. We implement \tool{} for C and Rust/Verus and evaluate it on 1,475 tasks. On
475 benchmarks from recent agentic verification tools, \tool{} solves 95.2\% of
the tasks. On 1,000 tasks from the SV-COMP 2026 ReachSafety track, averaging
1,427 LOC, \tool{} solves 91.5\% of the tasks, compared with 60.1\% for
CPAchecker. Ablations and comparison with a pure-agent baseline show that solver-aware
repair is effective and scalable.
\end{abstract}

\section{Introduction}

Large software projects continue to ship security-critical defects even after
extensive testing, review, and deployment. Empirical studies of mature
open-source systems show that vulnerabilities are distributed across real
project histories and related code regions, not merely isolated toy
examples~\cite{liu2020vulnerabilitydistribution}. Formal verification can
provide stronger assurance by proving program properties over all executions,
but applying it to realistic software still requires substantial proof
engineering. In deductive and SMT-backed verification workflows, the checker
reduces programs equipped with source-level specifications and auxiliary proof
artifacts to verification conditions (VCs). Proof engineers must supply loop
invariants, ghost state, intermediate assertions, and lemmas that expose enough
semantic structure for automated checkers to prove those
VCs~\cite{hahnle2019deductive,leino2010dafny}.

A common failure mode in this workflow is deceptively simple. A user, or an LLM
agent, adds a source-level specification that matches the intended
safety argument, yet the verifier still rejects the program. The reported
failure may be only a failed assertion, a timeout, or an unknown SMT result.
From this signal alone, it is unclear whether the specification is wrong,
whether an intermediate lemma is missing, whether the solver is distracted by
irrelevant context, or whether the obligation needs a different arithmetic,
bit-vector, quantified, or floating-point formulation. We call this property
\emph{solver dischargeability}: a program fact is not only intended to be true,
but is also expressed with the local facts, decomposition, and theory encoding
needed for the configured SMT backend to prove it.

\begin{figure}[t]
  \centering
  \setlength{\abovecaptionskip}{2pt}
  \includegraphics[width=\columnwidth]{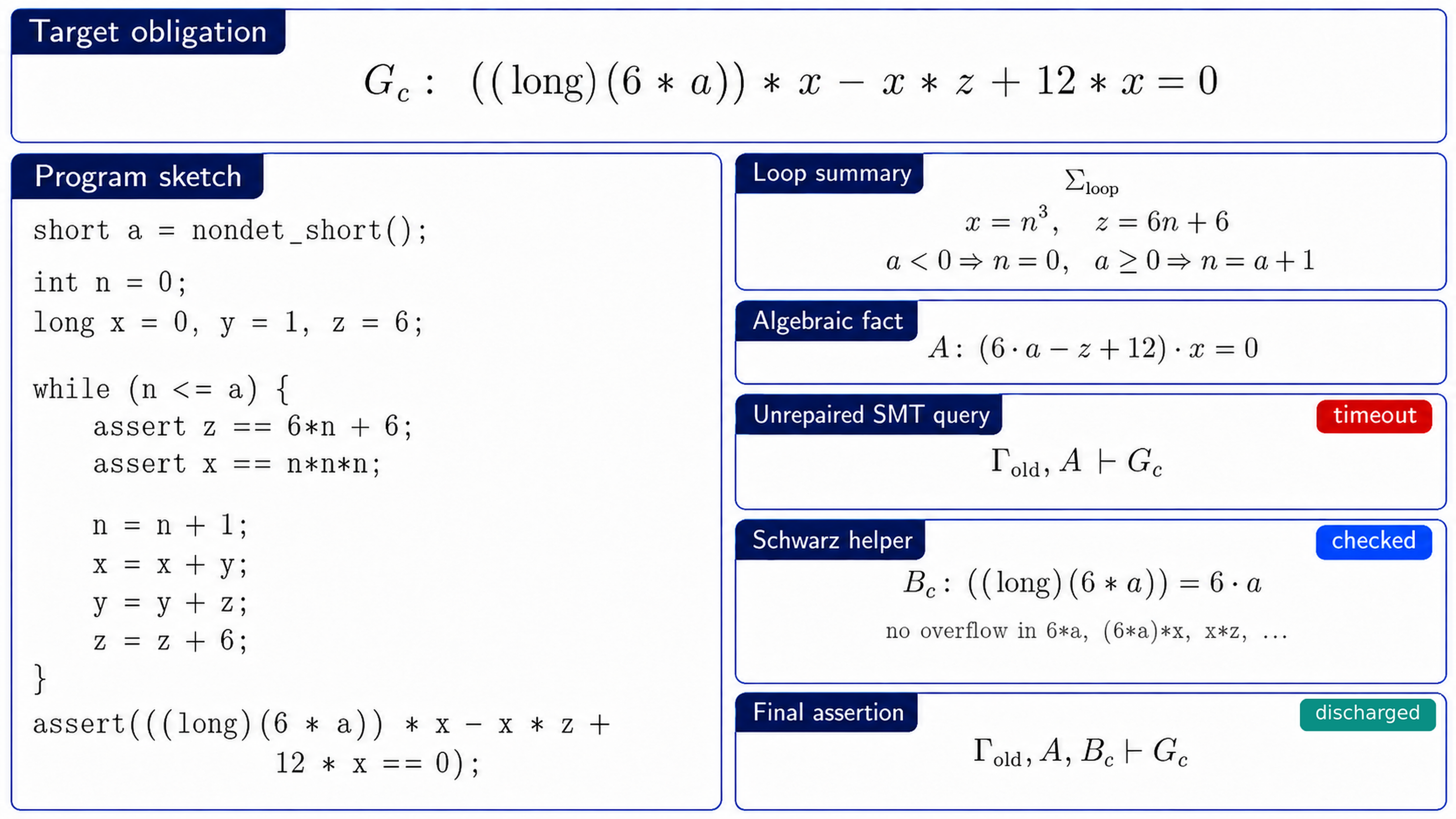}
  \caption{A C ReachSafety code fragment adapted from SV-COMP.}
  \label{fig:smt-repair-example}
\end{figure}

An example from the SV-COMP ReachSafety suite illustrates the gap. As shown in
Figure~\ref{fig:smt-repair-example}, the loop summary establishes
\(x=n^3\), \(z=6n+6\), and the final position of \(n\), from which the
algebraic fact \(A\) follows. The final assertion is a different SMT target
\(G_c\): since \(a\) is a \texttt{short}, the C expression \(6*a\) is first
evaluated under integer-promotion semantics, and the verifier must also prove
that the promoted multiplication and surrounding 64-bit arithmetic are defined.
The unrepaired query asks the old bit-vector context and \(A\) to prove
\(G_c\), and all configured solvers time out. The accepted \tool{} proof
introduces a local bridge \(B_c\) for the C-evaluation step: it checks that the
promoted term agrees with the 64-bit algebraic term and that the relevant
multiplications and additions do not overflow. With this bridge available, the
final assertion is discharged without forcing the top-level bit-vector query to
rediscover the C promotion and definedness reasoning.

Recent agentic-assisted verification systems seek to reduce the manual burden
of generating source-level specifications and auxiliary proof artifacts by
placing a large language model inside a checked repair loop. Systems built
around interactive theorem provers such as Rocq and Lean expose explicit goals,
hypotheses, tactics, and proof states checked by a trusted kernel
\cite{tu2026autorocq,zhao2026lemmanet}. This gives the model feedback for proof
search, but it also mixes the program's semantic intent with theorem-prover
context, tactic structure, and low-level axiomatic details. SMT-backed loops
offer a complementary route by reusing mature solvers and keeping the model
closer to source-level specifications, but current interfaces usually expose
only coarse verifier errors, timeouts, or unknown results
\cite{yang2025autoverus,liu2026kverus}. They therefore hide the local proof
context, missing helper facts, and theory-level choices that often determine
whether an SMT obligation can be discharged~\cite{leino2016triggers,
zhou2024contextpruning,barrett2025fixedsizebitvectors,jovanovic2017solving,
backeman2021interpolating}.

The key idea is to treat a failed proof not as a single request to rewrite the
whole specification, but as a diagnostic question: which local obligation
failed, what facts were visible there, and what checked helper fact or theory
view would make the obligation dischargeable? \tool{} realizes this idea with
three solver-facing optimization mechanisms. When the agent does not know where the
proof failed, checked snapshots expose the symbolic facts available at selected
program points. When the proof is too large for one solver step,
obligation-local lemmas let the agent propose a missing intermediate fact, which
must itself be proved before it can be used. When the solver is reasoning in an
unsuitable representation, theory-aware solver policies guide the agent toward
solver-friendly formulations for numeric, quantified, memory, and
floating-point obligations.

\tool{} does not ask the model to certify correctness. The model owns search:
it proposes source-level specifications, proof decompositions, local lemmas,
theory-aware solver policies, and counterexample inputs. The checker owns authority: it
preserves executable source code, regenerates proof obligations from trusted
front ends, checks SMT obligations with configured solvers, and accepts unsafe
claims only through concrete replay. Thus, every accepted safe or unsafe outcome
is grounded in deterministic checking rather than model rationale.

Our evaluation separates three questions that are often conflated in agentic
verification. First, can \tool{} reproduce or exceed the results of recent
agentic-verification systems on their own benchmark settings? Second, how much
do solver-facing lemmas and theory-aware policies contribute beyond source-level
specification generation alone? Third, what capability profile does \tool{}
exhibit on feature-rich C verification tasks? We evaluate
\tool{} on 1,475 tasks: 475
benchmarks from AutoRocq and KVerus, and a curated 1,000-task SV-COMP 2026 C
ReachSafety suite averaging 1,427 LOC. \tool{} solves 95.2\% of the 475 existing
agentic-verification tasks and 91.5\% of the SV-COMP tasks, with no incorrect
\tool{}-accepted verdicts. Comparisons with a pure-agent baseline and CPAchecker, the ReachSafety
champion, show that \tool{} improves the SV-COMP score from 72.5\% to 90.7\%
over the pure-agent baseline and achieves a higher accepted pass rate than
CPAchecker, 91.5\% versus 60.1\%.
Ablations show that local proof repair and theory-aware solver policies improve
agentic verification.

The paper makes three contributions.
\begin{itemize}
\item It identifies solver dischargeability as a bottleneck in SMT-backed
agentic verification, where plausible source-level specifications fail because
the repair loop does not expose missing local lemmas, irrelevant context, or
unsuitable theory formulations.
\item It presents \tool{}, a language-independent verification system that
turns failed verifier runs into checked, obligation-local repair tasks through
program-point snapshots, local lemmas, and a set of theory-aware solver
policies.
\item It evaluates \tool{} across 475 recent agentic-verification benchmarks
and a curated 1,000-task SV-COMP 2026 C ReachSafety suite, showing that
solver-aware repair is effective and scalable beyond small benchmarks.
\end{itemize}

\section{Background}

\begin{figure*}[t]
\centering
\includegraphics[width=0.92\textwidth]{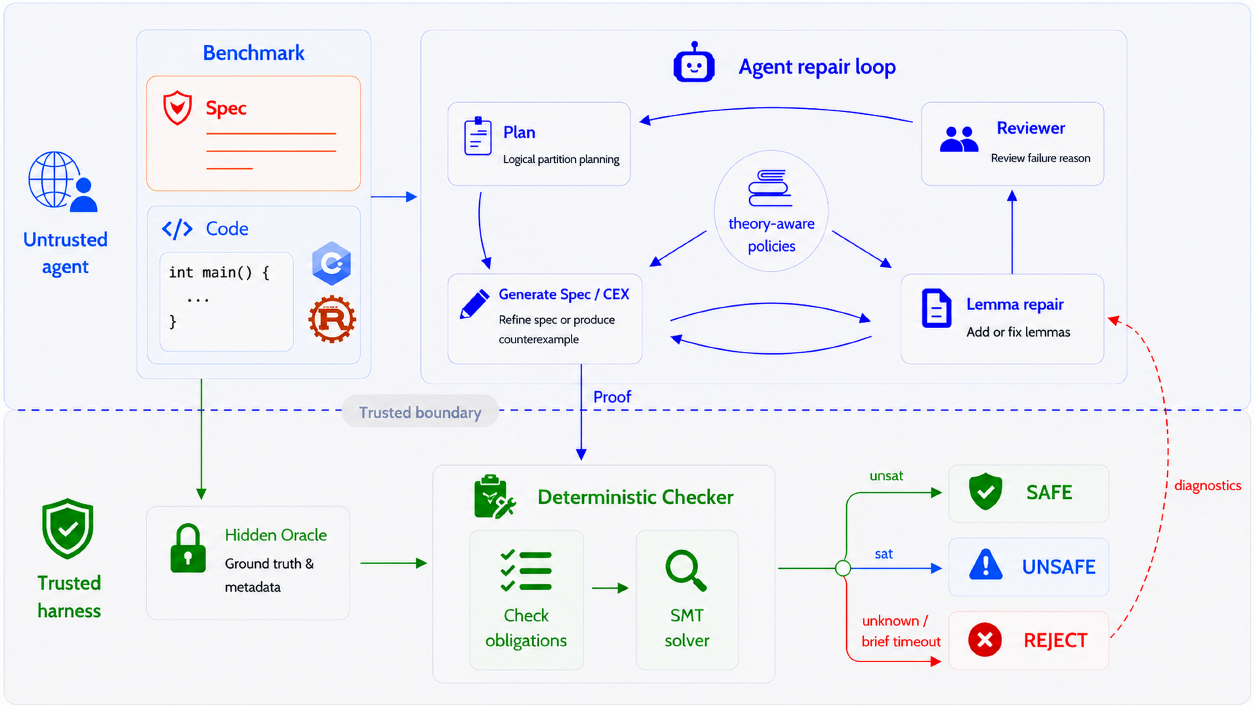}
\caption{Overview of \tool{}. The untrusted agent works above the trusted
boundary, where it plans, generates specifications or counterexamples, repairs
lemmas, and reviews failures under theory-aware policies. The trusted harness
keeps hidden metadata and accepts only results validated by deterministic
obligation checking, SMT solving, or concrete replay.}
\label{fig:overview-framework}
\end{figure*}

\subsection{Deductive Verification}

Deductive verification proves program properties by reducing source programs and
their specifications to logical proof obligations. The classical foundation is
Hoare-style reasoning: a triple \(\{P\}\ C\ \{Q\}\) states that command \(C\),
started in a state satisfying precondition \(P\), terminates only in states
satisfying postcondition \(Q\) if it terminates at all~\cite{floyd1967assigning,
hoare1969axiomatic}.
Weakest-precondition calculi and predicate transformers provide an algorithmic
way to compute sufficient preconditions for program constructs and are the basis
of many modern verification-condition generators~\cite{dijkstra1975guarded,
hahnle2019deductive}.

In practical tools, source-level specifications are the artifacts that
instantiate these Hoare-style predicates. Verification engineers write
preconditions and postconditions to state procedure contracts, assertions to
name intermediate predicates, loop invariants to summarize unbounded iteration,
frame conditions to bound side effects, and sometimes ghost state to expose
proof-only quantities. These specifications are not trusted as claims by
themselves. The verifier combines them with the program semantics and generates
checked verification conditions that justify each use of a specification:
assertions and postconditions must follow from the paths that reach them, loop
invariants generate initialization, preservation, and exit-use obligations, and
procedure calls are checked against their contracts and frame conditions. The
concrete encoding may use weakest preconditions, symbolic execution, SSA-like
intermediate forms, or a combination of these mechanisms, but the common
interface is a set of proof obligations that must be discharged before the
annotated program is accepted. Systems such as Frama-C/WP for ACSL-annotated
C~\cite{kirchner2015framac,baudin2008acsl}, Why3~\cite{filliatre2013why3}, and
Dafny~\cite{leino2010dafny} follow this contract-to-obligation pattern.

\subsection{SMT-backed Proof Discharge}

Many deductive verifiers discharge generated obligations with SMT solvers rather
than interactive proof. The front end encodes each obligation as a formula over
theories such as linear or nonlinear arithmetic, arrays, bit-vectors,
uninterpreted functions, and heap models. SMT-LIB provides a common language for
these formulas~\cite{barrett2010smtlib}, and solvers such as Z3 implement
decision procedures and heuristics for their combinations~\cite{demoura2008z3}.
To prove an obligation, the checker usually asks whether the negation of the
desired fact is unsatisfiable. If the solver returns \texttt{unsat}, the proof
step is accepted; if it returns \texttt{sat}, \texttt{unknown}, times out, or
the front end cannot translate the relevant source-language subset, the proof
step is rejected.

SMT automation is powerful, but its success depends on the chosen theory,
encoding, and query context. Quantified summaries rely on heuristic
instantiation, where trigger choices can create matching loops or unstable
verification times~\cite{leino2016triggers}. Arithmetic theories expose
different limitations. Linear real arithmetic has efficient simplex-style
procedures, but linear integer arithmetic must also enforce integrality and is
often handled with cuts, branch-and-bound, and layered
heuristics~\cite{dutertre2006fast,jovanovic2011cutting,griggio2010practical}.
Nonlinear integer arithmetic requires still more specialized
procedures~\cite{jovanovic2017solving}, while precise machine-integer reasoning
is often encoded with fixed-size bit-vectors~\cite{barrett2025fixedsizebitvectors}.
For bit-vectors, the dominant bit-blasting strategy can scale poorly as
bit-widths grow and arithmetic operators such as multiplication, division, and
remainder appear~\cite{zohar2022bitprecise,niemetz2024scalable,
chakraborty2017matching,teuber2020incremental}. Independently, irrelevant
assertions and assumptions in the query can make otherwise similar obligations
less stable or slower to discharge~\cite{zhou2024contextpruning}.

\subsection{Agent Harnesses}

Recent work defines an agent execution harness as the infrastructure around an
LLM agent and organizes it into seven layers: execution environment, tool
interface, context management, lifecycle and orchestration, observability,
verification, and governance~\cite{li2026agentharness}. We use this terminology
in a verification-specific sense. In \tool{}, the execution environment is the
isolated task workspace; the tool interface contains compilers, verifiers, SMT
solvers, and replay commands; context management consists of public task files,
snapshots, and diagnostic reports; lifecycle logic runs the checked repair loop;
observability records logs, SMT reports, and progress frontiers; verification is
the deterministic proof or counterexample checker; and governance is the
workspace access policy and acceptance rules.

\section{Overview}
\label{sec:overview}

Figure~\ref{fig:overview-framework} shows the \tool{} workflow and its trust
boundary. A benchmark exposes only the anonymized source program, target
specification, and compilation environment to the untrusted agent. Source-level
metadata such as original comments, ground-truth labels, and verifier-side
acceptance logic remain below the boundary inside the harness.

The agent side is organized as a repair loop rather than a one-shot generator.
The lead agent first makes a logical plan for the current proof attempt. It
then generates or refines source-level specifications, proposes a concrete
counterexample when the task appears unsafe, and invokes lemma repair when the
current proof attempt lacks an intermediate fact. A reviewer step inspects
failed attempts and routes the next iteration back to planning or repair. The
loop is guided by theory-aware policies, which tell the agent when an
obligation is likely to require a different arithmetic, bit-vector, memory, or
quantifier view, while leaving the soundness decision to the checker.

The checker side turns proposed artifacts into authoritative outcomes. For a
safe proof, the deterministic checker regenerates the relevant obligations from
the trusted source view and asks the configured SMT solver to discharge them; an
\texttt{unsat} result certifies that the checked obligation has no violating
model. For an unsafe answer, the harness accepts only a witness validated by
concrete replay. If the solver returns \texttt{unknown},
times out, or rejects a proposed lemma or specification, the attempt is not
accepted. Instead, the checker returns diagnostics that identify the failed
obligation or artifact. Timeout and \texttt{unknown} SMT outcomes are routed to
the lemma-repair stage, where the lead agent receives localized access to the
bottleneck SMT file and may edit it directly before the checker replays the
attempt. We next describe the technical design and core ideas behind these key
components.

\definecolor{KeyIntuitionBg}{HTML}{F6FAFF}
\definecolor{KeyIntuitionFrame}{HTML}{D7E4F5}
\newcommand{\keyintuitionbox}[1]{%
\par\vspace{0.45em}%
\noindent\begingroup
\setlength{\fboxsep}{5pt}%
\fcolorbox{KeyIntuitionFrame}{KeyIntuitionBg}{%
\begin{minipage}{\dimexpr\linewidth-2\fboxsep-2\fboxrule\relax}
\textbf{Key Intuition.} #1
\end{minipage}}%
\endgroup
\par\vspace{0.65em}}

\section{Technique}
\label{sec:technique}

A key intuition behind \tool{} is that a verification harness should minimize
two kinds of redundant context at the same time: redundant context presented to
the SMT solver, and redundant exploration performed by the LLM. If the harness
hides backend details behind high-level abstractions, the model sees only that
verification failed. It cannot tell whether the failure comes from a wrong
specification, missing context, or a solver-hostile encoding. In that setting,
repair often degenerates into
unfocused exploration. A second source of repetition is the lack of prior
knowledge about solver-friendly reductions: the model may repeatedly generate
formulations that are semantically equivalent to the desired proof fact but
inefficient for the theory combination being used.

\tool{} addresses these two issues by making solver feedback local and by
making recurring theory choices explicit. SMT-local lemmas expose failed solver
obligations as inspectable repair targets, while theory-aware policies encode
checked guidance about which annotation forms to use under different theory
scenarios. A
snapshot mechanism then keeps long proof attempts from accumulating irrelevant
state. The rest of this section describes these three components.

\subsection{SMT-Local Lemmas}

The smallest repair unit in \tool{} is an SMT-local lemma. For a source
location \(l\), the checker exposes a local obligation
\[
  O_l = \langle l, \Gamma_l, \varphi_l, \tau_l \rangle ,
\]
where \(\Gamma_l\) is the symbolic context visible at \(l\), \(\varphi_l\) is
the target fact, and \(\tau_l\) is the active theory view and proof policy. The
obligation is discharged when the solver can prove
\[
  \Gamma_l \models_{\tau_l} \varphi_l ,
\]
implemented as the unsatisfiability check
\(\mathsf{unsat}_{\tau_l}(\Gamma_l \wedge \neg \varphi_l)\). When this query
returns \texttt{unknown} or times out, the agent is not asked to rewrite the
entire proof attempt. Instead, the checker reports \(O_l\) and the generated
SMT file, so the agent can inspect the formula at the level where the solver
actually failed.

This feedback changes the shape of repair. The agent must decide whether the
SMT file is missing source semantics, carrying too much irrelevant context, or
using a theory view that does not match the intended fact. To test such
hypotheses, the agent may edit the local SMT file and invoke the solver
directly. It may also propose candidate lemmas \(m_1,\ldots,m_k\). The
diagnostic check first asks whether the strengthened local context is enough
for the target:
\[
  \Gamma_l \wedge m_1 \wedge \cdots \wedge m_k
  \models_{\tau_l} \varphi_l .
\]
If this query still fails, the proposed decomposition is insufficient. If it
succeeds, the checker turns back to the candidate lemmas and requires them to
be justified by the already checked context, for example by proving
\[
  \Gamma_l \wedge m_1 \wedge \cdots \wedge m_{i-1}
  \models_{\tau_l} m_i
  \quad\text{for each } i .
\]
Thus, an SMT-local lemma is a checked decomposition of the original obligation,
not an unchecked assumption. This whole hypothesis-testing loop runs directly
at the SMT level, avoiding repeated calls to the deterministic checker that
would regenerate and revalidate the proof attempt from the beginning. Direct
SMT edits are diagnostic rather than trusted proof evidence: a successful
experiment must be turned back into a source-level annotation that the checker
can regenerate and validate from the original program.

\tool{} also uses deliberately short default solver timeouts. In practice, when
the context is compact and the formula matches the intended theory, many
entailment steps that are obvious under a human mental model are discharged by
modern SMT solvers in milliseconds. Long default timeouts are therefore a poor
diagnostic signal: they delay the feedback loop and make it harder to tell
whether the issue is context size, theory choice, or a missing lemma. \tool{}
normally checks local obligations with aggressive timeouts, often in the
10--20 second range. The agent may selectively extend the timeout for an
obligation it believes is intrinsically harder, but the default behavior favors
fast failure and rapid hypothesis testing.

\keyintuitionbox{LLMs often fail at specification design because they cannot
see how a source-level specification is lowered into solver obligations. Exposing
and repairing SMT-local obligations removes the abstraction layers that would
otherwise cause repeated blind repair.}

\subsection{Theory-Aware Policies}

SMT-local feedback tells the agent where a proof attempt failed; theory-aware
policies tell it which formulation is likely to be efficient and sound for that
failure. These policies are not unchecked hints to the solver. They are scoped
proof directives exposed to the LLM as auxiliary context. Each policy specifies
when a source-level fact may be encoded in a particular SMT theory, which bridge
obligations must be generated, and when the front end must fall back to a more
precise theory. We illustrate this idea through the concrete structure of two
policies.

One policy concerns aggregate proofs over arrays and loops. A direct
specification may ask the agent to prove a final fact such as
\[
  s = \sum_{0 \leq k < n} A[k]
\]
or to show that one loop's accumulated elements are exactly canceled by a later
loop. Stating this as one global quantified formula mixes several theories and
program facts in a single obligation. Such a formula is often close to the
human proof idea but poor as a solver target and difficult for the LLM to repair
locally.

The aggregate policy instead tells the LLM to introduce proof-only ghost state
that advances with the program. For a prefix sum, the agent may maintain a
ghost prefix array \(P\):
\[
  P_{i+1} = \mathit{store}(P_i, i+1, \mathit{select}(P_i,i) + A[i]).
\]
The proof is then phrased as an inductive argument over local ghost updates,
with loop invariants or recursive-function summaries connecting each step to
the next. This form is usually closer to what SMT solvers can discharge than a
single aggregate formula, and it gives the LLM smaller obligations to inspect
and repair.

A second policy chooses between mathematical arithmetic and machine-integer
bit-vectors. \tool{} exposes theory selection as a first-class interface: the
LLM may choose the proof theory for each annotation, and the deterministic
checker ensures that the current formula can be lowered in that theory while
preserving the C semantics of the source program. This lets the agent use the
theory that matches the proof obligation instead of inheriting one global
encoding.

Encoding every scalar obligation as BitVec preserves C integer semantics, but
BitVec performs poorly on complex arithmetic, especially nonlinear products,
division, and modulo-style reasoning. For example, a simple interpolation proof
may need the arithmetic step
\[
  0 \leq t \leq d \;\wedge\; 0 \leq y_1-y_0
  \;\Rightarrow\;
  0 \leq t\cdot(y_1-y_0) \leq d\cdot(y_1-y_0).
\]
In an integer arithmetic view this is a small nonlinear fact; in BitVec it is a
bit-level statement about finite-width multiplication and wraparound, and the
solver often has little useful structure to exploit. Conversely, using
mathematical integers globally would be unsound when the program actually
depends on low-level machine behavior. The policy therefore tells the agent to
use LIA/NIA or their C-aware variants for value-level arithmetic, and to reserve
BitVec for facts whose truth depends on truncation, wraparound, or raw bit
layout.

These examples illustrate the role of the policy library. The LLM does not need
to rediscover, for every benchmark, that array aggregates should be represented
as scoped ghost summaries rather than monolithic quantified formulas, or that
BitVec is often the wrong default for scalar induction summaries. The policy
library captures these recurring solver-theory priors and turns them into
checked choices in the proof artifact.

\keyintuitionbox{Introducing prior knowledge about SMT solver theories reduces
redundant exploration caused by specifications that are semantically equivalent
but inefficient for the backend.}

\subsection{Snapshot Mechanism}

Snapshots keep the local repair loop from becoming global again. A snapshot is
a named boundary in the proof state: it records the program point, the checked
facts that remain in scope, the active proof policy, and any ghost or summary
state that has already been validated. Later proof steps can resume from this
boundary without replaying the entire exploration history or reintroducing
facts that are no longer relevant.

This mechanism is especially important for long programs and project-level
code. The path from the entry point to the current proof location may contain
substantial control flow and many already-checked proof steps. If every repair
attempt must replay that prefix, the feedback loop becomes dominated by
revalidation rather than by the local proof problem. Snapshots let the agent
stop at a previously checked point, keep only the compact state needed for the
next transition, and continue repair from there.

Snapshots are therefore an incremental engineering device, not the final
acceptance condition. After the proof pipeline reaches the end, \tool{} runs a
clean-gate check from the original source program. In that final pass, snapshots
are ignored as cached state, and all obligations are regenerated and checked
end to end.

\section{Implementation}
\label{sec:implementation}

We implemented \tool{} for C and Rust in about \(220\)K lines of non-test Rust
code. The implementation has three main components. First,
language-specific trusted front ends define how source-language semantics and
source-level annotations are lowered into verification conditions. Second, a
shared verification core checks formulas across several SMT theories and
solvers. Third, the harness layer provides the execution environment, including
Docker images and deterministic scripts for running the agent/checker loop.

Each language front end lowers the corresponding source semantics into formulas
over the shared verification core. The core then discharges the generated proof
obligations with an SMT solver bundle. Our evaluation environment uses
\textsc{cvc5}~1.3.1-dev.12.58cda4cdc, Bitwuzla~0.7.0-dev-main@d1f1bc2a, and
Z3~4.16.0. For short-timeout checks, \tool{} dispatches the obligation to the
configured solvers in parallel and uses their results to classify the obligation
as proved, refuted, unknown, or timed out.

For C, the trusted front end translates LLM-generated annotations together with
C semantic requirements into proof obligations. These requirements include, for
example, validity of array and pointer accesses and safety conditions for
dynamic \texttt{malloc}/\texttt{free} usage. Each obligation is checked under
the symbolic context available at the corresponding program point, typically as
an implication query encoded by asking whether the premises together with the
negated target are unsatisfiable. For SV-COMP ReachSafety tasks, the front end
also generates obligations requiring the path condition for reaching
\texttt{reach\_error()} to be unsatisfiable.

The C front end summarizes unbounded control flow in a standard deductive style.
Loop invariants and loop-exit summaries over-approximate the effects of loops,
while function contracts use \texttt{requires} and \texttt{ensures} clauses to
over-approximate function behavior. Therefore, when the checker proves a bad
state unreachable in this over-approximation, the conclusion is sound for the
executions represented by the trusted C semantics. The Rust front end follows
the same overall pattern: it lowers Rust programs and annotations into the
shared verification core, while modeling Rust-specific semantic details, such as
ownership, separately from the C memory model.

The harness layer wraps these trusted components in a deterministic execution
protocol. A deterministic script schedules the lead agent, receives the
agent-generated specification, invokes the backend checker, returns
trusted-side diagnostics, and resumes the agent for the next repair iteration.
The agent cannot inspect the trusted frontend lowering logic, checker internals,
or deterministic orchestration scripts. All input programs are anonymized before
they are shown to the agent, including program names, source comments, and extra
metadata, so the agent receives only the intended public program view.

\section{Evaluation}
\label{sec:evaluation}

The evaluation is designed to test how \tool{}'s framework design performs
across benchmarks of different scales and solver-theory demands.
We organize the study around three questions.

\noindent\textbf{RQ1: Effectiveness.} How does \tool{} perform compared with
existing traditional tools and agentic verification tools?

\noindent\textbf{RQ2: Solver-facing interface.} How much of the result comes
from obligation-local solver-facing guidance and theory-aware solver policies
rather than from source-level specification generation alone?

\noindent\textbf{RQ3: Capability boundaries.} On the 1,000-task C ReachSafety
suite, what capability profile does \tool{} exhibit across task features and
program scales, and what limitations remain in its current implementation?

\begin{table}[!t]
\centering
\caption{Benchmark suites used in the evaluation.}
\label{tab:prelim}
\footnotesize
\setlength{\tabcolsep}{3.1pt}
\begin{tabular}{@{}>{\raggedright\arraybackslash}p{0.39\columnwidth}
>{\raggedright\arraybackslash}p{0.25\columnwidth}cc@{}}
\toprule
Suite & Language & Tasks & Avg. LOC \\
\midrule
AutoRocq-ReachSafety & C/Rocq & 112 & 57 \\
AutoRocq-NoOverflow & C/Rocq & 50 & 79 \\
KVerus-File & Rust/Verus & 313 & 59 \\
VariousSVC-ReachSafety & C & 1000 & 1,427 \\
\midrule
Aggregate & Mixed & 1475 & 987 \\
\bottomrule
\end{tabular}
\end{table}

\begin{table}[!t]
\centering
\caption{LOC and dominant-feature distribution of the 1,000-task C
ReachSafety suite.}
\label{tab:svcomp-feature-loc-distribution}
\footnotesize
\setlength{\tabcolsep}{4.0pt}
\renewcommand{\arraystretch}{1.08}
\begin{tabular}{@{}lrrrrr@{}}
\toprule
Feature & $<100$ & 100--1000 & 1000--5000 & $>5000$ & Total \\
\midrule
Recursion & 120 & 7 & 0 & 0 & 127 \\
Control flow & 10 & 26 & 58 & 51 & 145 \\
Numeric & 140 & 81 & 68 & 18 & 307 \\
Floating point & 114 & 100 & 22 & 22 & 258 \\
Array/pointer & 77 & 83 & 3 & 0 & 163 \\
\midrule
Total & 461 & 297 & 151 & 91 & 1000 \\
\bottomrule
\end{tabular}
\end{table}

\newcommand{\evalNA}{\textcolor{black!45}{--}}
\newcommand{\evalBest}[1]{\textbf{#1}}
\newcommand{\evalHeader}[1]{\leavevmode\raisebox{-0.92\baselineskip}[0pt][0pt]{#1}}
\newcommand{\evalKVerus}{KVerus\textsuperscript{\dag}}
\definecolor{RQAnswerBg}{HTML}{F7FAF6}
\definecolor{RQAnswerFrame}{HTML}{B7C8B4}
\newcommand{\rqanswerbox}[2]{%
\par\vspace{0.55em}%
\noindent\begingroup
\setlength{\fboxsep}{5.5pt}%
\setlength{\fboxrule}{0.65pt}%
\fcolorbox{RQAnswerFrame}{RQAnswerBg}{%
\begin{minipage}{\dimexpr\linewidth-2\fboxsep-2\fboxrule\relax}
\textbf{Answer to #1.} #2
\end{minipage}}%
\endgroup
\par\vspace{0.65em}}

\begin{table*}[!t]
\centering
\caption{Cross-benchmark comparison.}
\label{tab:cross-benchmark-comparison}
\footnotesize
\setlength{\tabcolsep}{4.2pt}
\renewcommand{\arraystretch}{1.16}
\begin{tabular}{@{}p{0.17\textwidth}p{0.078\textwidth}
>{\centering\arraybackslash}p{0.043\textwidth}
@{\hspace{0.026\textwidth}}>{\centering\arraybackslash}p{0.098\textwidth}
>{\centering\arraybackslash}p{0.098\textwidth}
>{\centering\arraybackslash}p{0.098\textwidth}
>{\centering\arraybackslash}p{0.098\textwidth}
>{\centering\arraybackslash}p{0.098\textwidth}@{}}
\toprule
\evalHeader{Benchmark suite} &
\evalHeader{Language} &
\evalHeader{Tasks} &
\multicolumn{5}{c@{}}{Direct passes (pass rate)} \\
\cmidrule(l){4-8}
 & & & \cellcolor{black!7}\tool{} & AutoRocq & AutoVerus &
\evalKVerus{} & CPAchecker \\
\midrule
AutoRocq-ReachSafety & C/Rocq & 112 &
\cellcolor{black!7}\evalBest{106 (94.6\%)} & 45 (40.2\%) & \evalNA{} & \evalNA{} & 25 (22.3\%) \\
AutoRocq-NoOverflow & C/Rocq & 50 &
\cellcolor{black!7}\evalBest{47 (94.0\%)} & 10 (20.0\%) & \evalNA{} & \evalNA{} & 45 (90.0\%) \\
KVerus-File & Rust/Verus & 313 &
\cellcolor{black!7}\evalBest{299 (95.5\%)} & \evalNA{} & 217 (69.3\%) & 251 (80.2\%) & \evalNA{} \\
VariousSVC-ReachSafety & C & 1000 &
\cellcolor{black!7}\evalBest{915 (91.5\%)} & \evalNA{} & \evalNA{} & \evalNA{} & 601 (60.1\%) \\
\midrule
Aggregate & Mixed & 1475 &
\cellcolor{black!7}\evalBest{1367 (92.7\%)} & \evalNA{} & \evalNA{} &
\evalNA{} & \evalNA{} \\
\bottomrule
\end{tabular}
\vspace{0.35em}

\raggedright
\scriptsize
\emph{Note:} Result cells report direct passes, with pass rates in parentheses;
a dash indicates that the tool does not target the suite or no comparable
result is available. AutoRocq results are mapped from VC-level outcomes to
property-specific source-program outcomes: a source benchmark is counted solved
only when all generated VCs in the corresponding property slice are solved and
rechecked. In AutoRocq-ReachSafety, the CPAchecker false-unsafe result is
counted unsolved.
\textsuperscript{\dag}\,KVerus was not open-sourced at evaluation time, so we
report the paper's Claude Sonnet 4.0 result.
\end{table*}

\subsection{Evaluation Setting}

Table~\ref{tab:prelim} summarizes the benchmark suites used in the evaluation.
The evaluation uses two language settings. The C portion combines AutoRocq's
SV-COMP-derived suites with a curated 1,000-task SV-COMP 2026 C ReachSafety
suite. The AutoRocq C suites contain 112 ReachSafety programs and 50 NoOverflow
programs whose Rocq obligations are generated by Frama-C/WP through
Why3~\cite{tu2026autorocq,kirchner2015framac,filliatre2013why3}. The
specifications used for those obligations are generated as part of AutoRocq's
benchmark construction; when running \tool{}, we remove this metadata and expose
only the anonymized source program, target property, and compilation
environment. The curated C suite is selected from the SV-COMP 2026 ReachSafety
benchmark set~\cite{beyer2025svcomp,svcomp2026results}; it is stratified by
feature category and LOC bucket and covers structured loops, arrays and
pointers, floating-point and numeric reasoning, bit-level operations, and
recursion. Table~\ref{tab:svcomp-feature-loc-distribution} reports the
resulting LOC-by-feature distribution for the 1,000 tasks. The benchmark
excludes cases containing cyclic \texttt{goto}, which are outside the current
frontend scope.

The Rust/Verus portion uses 313 single-file proof tasks from the
AutoVerus/KVerus setting~\cite{yang2025autoverus,liu2026kverus}. Some original
tasks include synthesis-generated proof helpers. We remove all such helper
proof material and keep only the target to be verified and the generated
program as the input to \tool{}.

Each task is imported into the two-view case format from
Section~\ref{sec:overview}. \tool{} is not given separate safe and unsafe modes:
each input may be safe or unsafe, and the prompt contains no hint favoring
either outcome. We run all tools with a 4-hour wall-clock timeout per task. For
agentic tools, we do not cap the number of model calls or repair rounds within
that timeout. Each task is limited to at most 8 CPU cores and 12~GB of memory.
All experiments run on a machine with an AMD Ryzen 9 5950X 16-core CPU
(32 hardware threads) and 128~GB RAM. We use Codex CLI 0.130.0 with
GPT-5.5 xhigh as the underlying lead agent for \tool{} and for the pure-agent
baseline. We report direct passes as the primary metric because they correspond
to outcomes accepted by deterministic infrastructure rather than by model
judgment.

\subsection{RQ1: Effectiveness}

\tool{} is compared with four representative tools:
\begin{itemize}
\item \textbf{AutoRocq} is an agentic verification tool for proving program-verification
obligations exported from Frama-C/WP and Why3 into Rocq~\cite{tu2026autorocq}.
It uses Rocq proof states and kernel-checked proof scripts as feedback for
iteratively constructing proofs for the generated verification conditions.
\item \textbf{AutoVerus} targets Rust programs written for the Verus verifier
and uses LLM agents to synthesize Verus specifications and proof
code~\cite{yang2025autoverus}. It organizes proof generation into preliminary
construction, verifier-guided refinement, and debugging from Verus errors.
\item \textbf{KVerus} builds on Verus and combines dependency-graph analysis,
semantic lemma generation and retrieval, toolchain knowledge, and error-driven
self-refinement~\cite{liu2026kverus}. It uses dependency structure and reusable
proof knowledge as persistent context for project-level proof repair.
\item \textbf{CPAchecker} is a traditional software verifier for C based on
configurable program analysis, integrating multiple abstract domains and
model-checking strategies under a common framework~\cite{beyer2009cpachecker}.
It ranked first in the C.ReachSafety category of SV-COMP
2026~\cite{svcomp2026results}.
\end{itemize}

Table~\ref{tab:cross-benchmark-comparison} is the main comparison table.
AutoRocq proves generated Rocq VCs, AutoVerus and KVerus synthesize Verus proof
artifacts, and CPAchecker checks C reachability or overflow
directly~\cite{beyer2009cpachecker}. \tool{} is evaluated through different
language front ends and the same hidden harness across the suites.

\tool{} produces 1367
direct passes across 1475 tasks (92.7\%). On the 475 tasks drawn from recent
agentic verification benchmarks, it passes 452 tasks (95.2\%), including
106/112 AutoRocq ReachSafety tasks, 47/50 AutoRocq NoOverflow tasks, and
299/313 KVerus-File tasks. On the 1,000-task SV-COMP 2026 ReachSafety suite,
\tool{} passes 915 tasks (91.5\%), compared with 601 (60.1\%) for
CPAchecker.
The largest gains appear on the AutoRocq ReachSafety and KVerus-File suites,
where the existing tools operate through generated Rocq or Verus proof
interfaces, which can be affected by tool abstraction layers and large
amounts of redundant proof context. By contrast, \tool{} keeps repair tied to
source-level annotations and local solver-facing obligations.

\begin{figure}[!t]
\centering
\includegraphics[width=\columnwidth]{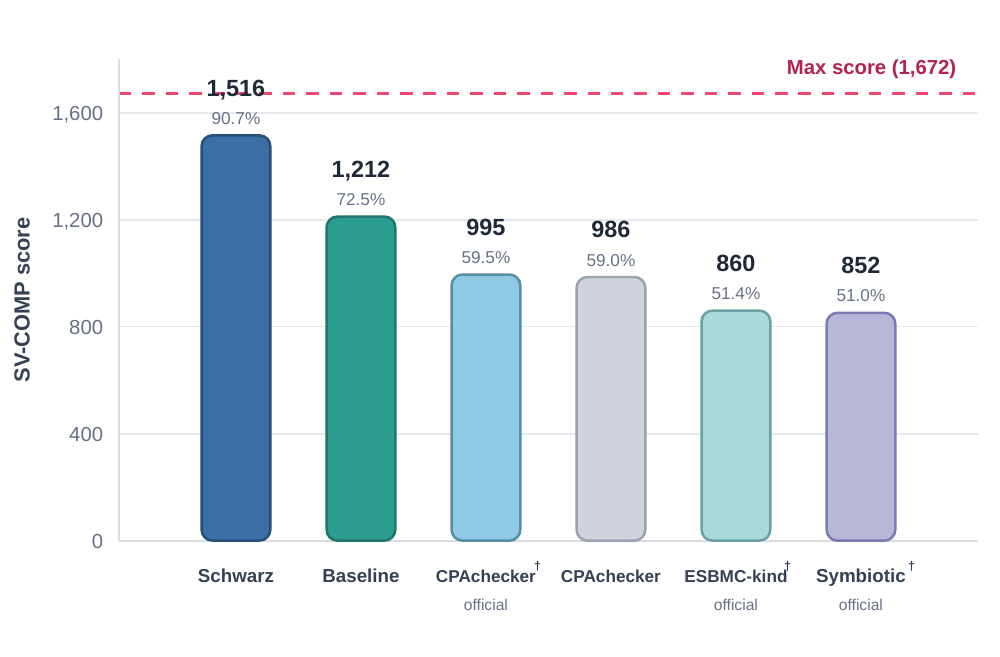}
\caption{SV-COMP-style scores on the 1,000-task SV-COMP 2026 ReachSafety
suite. Daggered bars are official SV-COMP 2026 per-task results projected to
the same local subset.}
\label{fig:svcomp-score-bars}
\end{figure}

\begin{table*}[!t]
\centering
\caption{Ablation comparison for RQ2.}
\label{tab:rq2-ablation}
\footnotesize
\setlength{\tabcolsep}{3.0pt}
\renewcommand{\arraystretch}{1.16}
\begin{tabular}{@{}p{0.185\textwidth}
>{\raggedright\arraybackslash}p{0.075\textwidth}
>{\centering\arraybackslash}p{0.045\textwidth}
>{\centering\arraybackslash}p{0.100\textwidth}
@{\hspace{0.012\textwidth}}>{\centering\arraybackslash}p{0.115\textwidth}
>{\centering\arraybackslash}p{0.165\textwidth}
>{\centering\arraybackslash}p{0.155\textwidth}@{}}
\toprule
\evalHeader{Benchmark suite} &
\evalHeader{Language} &
\evalHeader{Tasks} &
\multicolumn{1}{c}{Full} &
\multicolumn{3}{c@{}}{Ablation} \\
\cmidrule(lr){4-4}\cmidrule(l){5-7}
 & & & \tool{} & No Policies & No SMT-Local Lemmas & No Policies/Lemmas \\
\midrule
AutoRocq-ReachSafety & C/Rocq & 112 & \cellcolor{black!7}106 (94.6\%) & 104 (92.9\%) &
91 (81.3\%) & 82 (73.2\%) \\
AutoRocq-NoOverflow & C/Rocq & 50 & \cellcolor{black!7}47 (94.0\%) & 45 (90.0\%) &
43 (86.0\%) & 46 (92.0\%) \\
KVerus-File & Rust/Verus & 313 & \cellcolor{black!7}299 (95.5\%) & 281 (89.8\%) &
256 (81.8\%) & 230 (73.5\%) \\
VariousSVC-ReachSafety & C & 1000 & \cellcolor{black!7}915 (91.5\%) & 871 (87.1\%) &
756 (75.6\%) & 723 (72.3\%) \\
\midrule
Aggregate & Mixed & 1475 & \cellcolor{black!7}\evalBest{1367 (92.7\%)} & 1301 (88.2\%) &
1146 (77.7\%) & 1081 (73.3\%) \\
\bottomrule
\end{tabular}
\end{table*}

Figure~\ref{fig:svcomp-score-bars} compares tools on the 1,000-task
SV-COMP 2026 ReachSafety suite using the SV-COMP scoring rule: a correct safe
answer receives two points, a correct unsafe answer receives one point, a wrong
safe answer receives $-32$ points, and a wrong unsafe answer receives $-16$
points. The ``Baseline'' bar
is the pure-agent baseline: the same underlying agent is asked to classify the
program's safety property as safe or unsafe without any formal assistance.
The base model maintains a high solve rate without any formal assistance, but
it is still severely affected by hallucination: on the 1,000 tasks, it emits
17 wrong labels, including 10 wrong safe answers and 7 wrong unsafe answers.
Two additional expected-safe tasks receive no scored answer. Under the SV-COMP
penalty rule, these outcomes reduce its score to
1212/1672 (72.5\%). \tool{} scores 1516/1672 (90.7\%) with 915 accepted passes:
601 safe proofs and 314 unsafe replays. It therefore substantially improves the
score over both the pure-agent baseline and CPAchecker
(986/1672, 59.0\%), while retaining deterministic checking for every accepted
outcome.

\noindent\textbf{Failure Analysis.}
We analyze the remaining cases qualitatively rather than reporting a separate
non-pass taxonomy. The dominant source of missed passes is missing semantic
coverage in the current trusted front ends and solver backends. Examples
include standard-library or libm functions whose precise semantics are not yet
modeled, such as \texttt{tanf}, \texttt{sqrtf}, and string routines such as
\texttt{strncmp}; C constructs that require finer bit-level layout modeling,
especially union-heavy code; and cases where the target location requires a
large number of source-level specifications before the agent can establish a
sufficient symbolic context. These implementation and backend semantic gaps
account for roughly 80\% of the unresolved cases in our logs. By contrast,
plain time exhaustion is less common: in the 1,000-task SV-COMP suite, only 13
tasks remain unresolved primarily because the run ran out of time before the
needed specification route could be generated and checked.

\rqanswerbox{RQ1}{\tool{} solves 1367/1475 tasks across the evaluation suite
(92.7\%), including 452/475 tasks from recent agentic-verification benchmarks
and 915/1000 SV-COMP ReachSafety tasks. On the SV-COMP subset, it improves the
score from 1212/1672 for the pure-agent baseline to 1516/1672, while also
exceeding CPAchecker's accepted pass rate, 91.5\% versus 60.1\%.}

\subsection{RQ2: Solver-Facing Interface}

RQ2 tests the contribution of local solver-facing optimization, not merely
whether the LLM can produce more source-level specifications. Table~\ref{tab:rq2-ablation}
uses a two-factor ablation: theory-aware solver policies are either enabled or
disabled, and SMT-local lemmas are either enabled or disabled. This isolates
whether the gains come from choosing solver-friendly theory views, from adding
checked intermediate proof steps, or from their combination.

Table~\ref{tab:rq2-ablation} shows that both components contribute to the
overall result, but they do so at different magnitudes. With the full
solver-facing interface, \tool{} passes 1367/1475 tasks (92.7\%). Removing
theory-aware policies reduces the aggregate result to 1301 passes (88.2\%),
a loss of 66 tasks. Removing SMT-local lemmas has a larger effect: the result
drops to 1146 passes (77.7\%), losing 221 tasks. When both components are
disabled, the system passes 1081 tasks (73.3\%), 286 fewer than the full
configuration.

\begin{figure}[!t]
\centering
\includegraphics[width=\columnwidth]{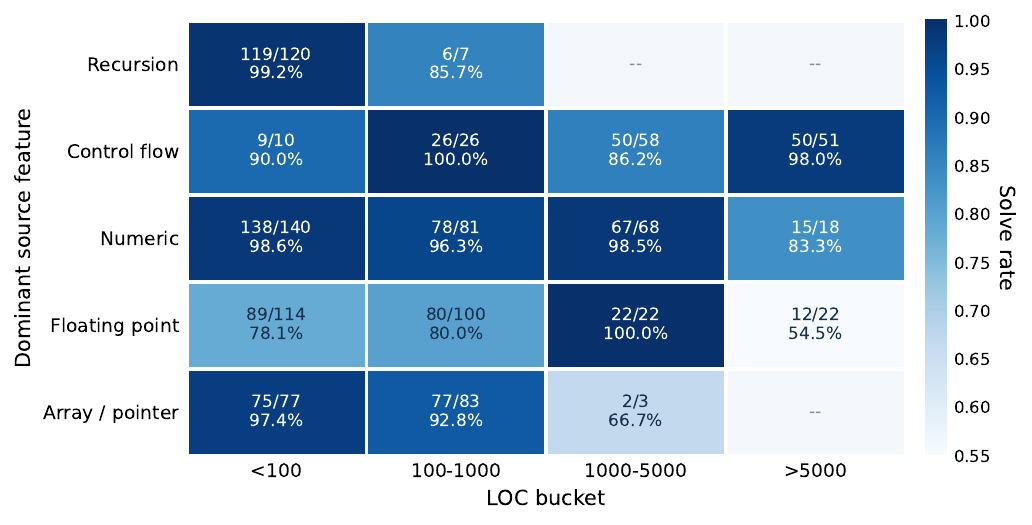}
\caption{\tool{} accepted passes by dominant source feature and LOC bucket on
the 1,000-task SV-COMP 2026 ReachSafety suite. Each cell reports accepted
passes over tasks in that bucket and the corresponding pass rate.}
\label{fig:svcomp-feature-heatmap}
\end{figure}

The largest and most stable drop comes from disabling SMT-local lemmas. On the
1,000-task C ReachSafety suite, the pass rate decreases from 91.5\% to 75.6\%;
on KVerus-File, it decreases from 95.5\% to 81.8\%; and on
AutoRocq-ReachSafety, it decreases from 94.6\% to 81.3\%. These drops support
the design premise of \tool{}: many failed obligations are not solved by simply
adding more source-level annotations. The agent often needs a way to expose a
small missing proof step, test it directly against the generated SMT
obligation, and then lift the successful step back into a checker-accepted
source annotation. Without this local repair loop, the agent must infer the
solver bottleneck indirectly from a larger verifier failure, which leads to
more redundant exploration.

Theory-aware policies provide a smaller but still consistent benefit on the
larger suites. Removing policies loses 44 passes on VariousSVC-ReachSafety and
18 passes on KVerus-File. This is expected: policies do not replace proof
construction, but they bias the agent toward annotation forms that fit the
dominant theory view of the obligation. Their effect is therefore most visible
when the task mixes arrays, machine integers, bit-level facts, and arithmetic
summaries, where several semantically equivalent annotations can produce very
different solver behavior. The AutoRocq-NoOverflow suite is less diagnostic:
all configurations remain within four tasks of the full system, suggesting
that these short arithmetic obligations are already close to saturation under
the current harness.

The combined ablation is worse than either single ablation on three of the four
suites and on the aggregate. This indicates that the two mechanisms are
complementary. SMT-local lemmas give the agent a checked way to decompose a
hard obligation; policies help it choose a decomposition that the solver can
handle efficiently. The gap between the full system and the combined ablation
is especially large on KVerus-File (299 versus 230 passes) and on
VariousSVC-ReachSafety (915 versus 723 passes), where proof search often
depends on both local intermediate facts and theory-sensitive annotation
choices.

\rqanswerbox{RQ2}{The solver-facing interface accounts for a substantial
fraction of \tool{}'s performance. SMT-local lemmas are the dominant
contributor, and theory-aware policies add further gains on feature-rich tasks;
together, they raise the aggregate pass rate from 73.3\% to 92.7\%.}

\subsection{RQ3: Capability Boundaries}

RQ3 uses the 1,000-task C ReachSafety suite to characterize the boundary of
\tool{}'s current capability. Figure~\ref{fig:svcomp-feature-heatmap} breaks
the 915 accepted passes down by dominant source feature and LOC bucket. The
main observation is that the pass-only metric remains high across several
different kinds of tasks, not only on short programs: \tool{} solves 119/120
recursion cases and 138/140 numeric cases below 100 LOC, 77/83 array-pointer
cases and all 26 control-flow cases in the 100--1000 LOC bucket, 67/68 numeric
cases in the 1000--5000 LOC bucket, and 50/51 control-flow-heavy cases above
5000 LOC. These results are consistent with the role of local solver feedback,
theory policies, and snapshots beyond the smaller AutoRocq-derived C tasks.

The heatmap also exposes the current boundary. Floating-point tasks are the
least stable region: \tool{} solves 89/114 tasks below 100 LOC, 80/100 tasks in
the 100--1000 LOC bucket, and 12/22 tasks above 5000 LOC, even though it solves
all 22 floating-point tasks in the 1000--5000 LOC bucket. Manual inspection
indicates that many of these
misses are caused by missing built-in models for library and libm behavior,
including functions such as \texttt{sqrtf} and \texttt{tanf}, rather than by
the absence of a high-level proof strategy. A second boundary appears in
longer proof routes where the target location is preceded by many semantically
relevant checks and updates. Snapshots reduce the cost of replaying already
proved prefixes, but more aggressive snapshot placement and context
summarization would expose a smaller checked state near the target during
repair iterations. Finally, a small number of memory-layout
cases still require more precise modeling of layout-sensitive C constructs,
especially unions and bit-level reinterpretation.

\rqanswerbox{RQ3}{\tool{} is strongest on integer and numeric reasoning,
recursion, and large control-flow tasks where source-level summaries can be
checked locally and reused through snapshots. Its current limitations are
concentrated in floating-point and library semantics, layout-sensitive C
features, and long proof routes that require many specifications before the
target location.}

\subsection{Threats to Validity}

The evaluation has two main threats. First, benchmark suites differ in task
granularity and proof artifact: Rocq VCs, Verus proof artifacts, and C source
programs are not interchangeable units. We therefore report per-suite results
instead of treating the aggregate as a universal leaderboard. Second, the
overall results may be affected by nondeterminism in model outputs and by the
incomplete coverage of the current prototype. A specification may admit
multiple proof routes and theory choices, while some programs fall into
semantic regions that \tool{} does not yet fully support, leading to blocked
runs rather than accepted outcomes.

\section{Related Work}

\newcommand{\rwtopic}[1]{\smallskip\noindent{\bfseries #1.}\ }

\rwtopic{Agentic program verification}
The closest line of work uses language-model agents to generate or repair proof
artifacts around a verifier. AutoRocq proves program-verification obligations
exported from Frama-C/WP and Why3 into Rocq, giving the agent explicit theorem
prover states and a trusted kernel~\cite{tu2026autorocq,kirchner2015framac,
filliatre2013why3}. LemmaNet further adds source-guided auxiliary lemma
generation at the Rocq level: it maps the semantics of source-level helper
specifications into Rocq lemmas and improves the solve rate over the AutoRocq
pipeline~\cite{zhao2026lemmanet}.

This Rocq-based route has a different cost profile from SMT-backed deductive
repair. As illustrated by LemmaNet's Fig.~2, the Frama-C/Why3-to-Rocq path must
compile program semantics into Rocq formulas and often introduces verbose
contexts and axiomatic encodings around the target lemma. When an early lemma in
the proof route needs to be reconstructed, Rocq compilation and proof replay can
become a bottleneck in the agentic repair loop, especially on larger programs.
\tool{} takes the complementary design: it repairs source-level annotations while
using short-timeout SMT obligations as the local feedback unit. This avoids
replaying a large Rocq development for each hypothesis test and lets the agent
reuse mature SMT theory reasoning, lemmas, and solver heuristics instead of
reproving facts already modeled by the solver. The interactive theorem-proving
route remains valuable for domains where SMT support is weak, such as SQL-style
bag reasoning with aggregation and order.

AutoVerus and KVerus are agentic frameworks built on the Rust Verus
verifier~\cite{yang2025autoverus,liu2026kverus}. AutoVerus introduces a simple
repair loop that iteratively fixes generated proofs, while KVerus targets
project-level proof generation by using static program information to help the
LLM understand interprocedural data and control dependencies. Both systems also
use proof lemmas to improve the verification of complex specifications. Relative
to \tool{}, however, their lemmas remain at the Verus frontend level: they help
the model repair Verus-facing specifications, but they do not directly explain
why the SMT obligations produced after Verus lowering return \texttt{unknown}
or time out. Such solver-level failures are common in larger verification
tasks, and modeling them explicitly is the core problem addressed by \tool{}.

Earlier LLM-assisted formal-verification work is broader than these verifier
agents. One line asks LLMs to predict invariants or proof hints and then checks
or ranks the candidates with symbolic tools~\cite{pei2023invariants,
kamath2023inductive,chakaraborty2023ranking,wu2024lam4inv,wei2026quokka,
bhat2024synver}. Another line targets proof assistants and formal theorem
proving: DSP uses informal proof drafts to guide formal proof sketches, Baldur
generates and repairs whole Isabelle proofs, LeanDojo/ReProver augments Lean
proving with premise retrieval, and COPRA performs stateful in-context tactic
search for Lean and Coq~\cite{jiang2023dsp,first2023baldur,yang2023leandojo,
thakur2024copra}. These systems established the checked-generation pattern:
the model proposes logical candidates, and a deterministic checker accepts or
rejects them. \tool{} builds on this pattern but studies a deeper repair
problem:
how to expose local SMT failures, theory choices, and replay boundaries in a
source-level program-verification harness.

\rwtopic{LLM-assisted SMT reasoning}
Recent work has begun to use LLMs to improve SMT and constraint solving more
directly. AquaForte uses LLM-generated semantic guidance to instantiate
quantified formulas with uninterpreted functions, reducing the search space for
traditional SMT solvers~\cite{lv2026aquaforte}. Another line uses LLMs to
generate auxiliary lemmas for constraints with inductive or recursive
definitions, and then asks symbolic solvers to check the validity and usefulness
of those lemmas~\cite{feng2026llminductive}. These systems are close in spirit
to \tool{} because they treat LLM output as hypotheses that must be checked by a
solver.

The main difference is scope and information source. Existing LLM-assisted SMT
systems are designed around particular theory fragments, such as quantified
uninterpreted functions or recursive definitions. \tool{} instead organizes
solver-facing repair as part of a program-verification harness, combining local
lemmas, snapshots, and theory policies under a single checked workflow. Moreover,
after source programs have been lowered into SMT, part of the source-level
semantic structure has already been abstracted away. \tool{} generates and
repairs lemmas using both the SMT obligation and the corresponding source-level
program context, while approaches that optimize SMT formulas alone must work
without that additional semantic view. These directions are therefore largely
orthogonal: LLM-guided SMT techniques optimize general solver tasks, while
\tool{} combines similar solver-facing repair with the source-level semantics of
program verification, making the optimization more effective for verification
tasks.

\rwtopic{Traditional verification tools and specification languages}
\tool{} is complementary to mature verification tools. CPAchecker provides a
configurable framework for C software verification and a strong SV-COMP
baseline~\cite{beyer2009cpachecker,beyer2025svcomp}. CBMC performs
bit-precise bounded model checking for C assertions~\cite{kroening2023cbmc},
and KLEE uses symbolic execution to generate high-coverage tests for systems
programs~\cite{cadar2008klee}. Frama-C/WP generates weakest-precondition VCs
from ACSL-specified C programs~\cite{kirchner2015framac,baudin2008acsl}, while
Dafny and Verus provide SMT-backed specification and proof systems for
high-level or Rust-like programming models~\cite{leino2010dafny,dafnyref,
yang2025autoverus}. \tool{} does not replace these systems with model judgment.
It studies how an agentic harness can expose enough proof interface for an LLM
to construct useful source-level specifications and proof facts while leaving
correctness judgments to deterministic checkers.

\section{Discussion}
\label{sec:discussion}

During the development of \tool{}, we observed that the plan-execute-review
pattern has become a relatively standard outer loop for agentic systems. As this
outer structure stabilizes, the main question is no longer whether a harness can
ask an agent to plan, act, and revise, but what information and control surfaces
the harness exposes inside that loop. The design of \tool{} follows this view:
its purpose is not to show the agent the front-end lowering code and ask it to
infer the verifier's behavior. Instead, \tool{} exposes solver-level context and
the correspondence between source-level specifications, symbolic contexts, and
SMT obligations. This makes the logical path from source annotations to SMT
queries explicit enough for the agent to diagnose local proof failures, while
keeping the trusted lowering and checking logic outside the agent's control.

We also found that domain knowledge is becoming an essential part of harness
design. Many useful verification techniques are not simply generic reasoning
patterns that a base model can be expected to know reliably. They include recent
research results, solver-specific engineering heuristics, and low-level
improvements in verification tools, all of which may appear too late or too
sparsely to be absorbed into the base model during pretraining. An effective
harness therefore needs mechanisms for bringing this knowledge into the repair
loop. One route is to let the harness accumulate reusable experience from prior
runs as long-term context. Another is to encode expert knowledge directly, as
\tool{} does with theory-aware policies that guide the agent toward
solver-friendly annotations. In both cases, the harness is not merely a wrapper
around an LLM; it becomes the place where domain expertise is organized,
checked, and made operational.

\section{Conclusion}

\tool{} studies agentic verification as a harness problem: the model searches
over annotations, decompositions, and theory choices, while correctness remains
with deterministic front ends, SMT checking, and concrete replay. Its central
lesson is that plausible source-level specifications are not enough; the harness
must expose the symbolic context, missing proof steps, and theory choices that
make SMT obligations dischargeable. Across 1,475 tasks, \tool{} solves 1367
cases, including 95.2\% of recent agentic-verification benchmarks and 91.5\% of
the 1,000-task SV-COMP 2026 ReachSafety suite. The remaining failures are mostly
semantic coverage gaps, suggesting that progress depends on harnesses that
organize domain knowledge and expose checked solver-facing repair to agents.

\smallskip
\noindent\textbf{Use of Generative AI.} The authors used generative AI tools to
assist in generating Figures~\ref{fig:smt-repair-example}
and~\ref{fig:overview-framework}; all figure content was manually checked.

\bibliographystyle{IEEEtran}
\bibliography{references}

@article{kamath2023inductive,
  author = {Adharsh Kamath and Aditya Senthilnathan and Saikat Chakraborty and Pantazis Deligiannis and Shuvendu K. Lahiri and Akash Lal and Aseem Rastogi and Subhajit Roy and Rahul Sharma},
  title = {Finding Inductive Loop Invariants using Large Language Models},
  journal = {arXiv preprint arXiv:2311.07948},
  year = {2023},
  doi = {10.48550/arXiv.2311.07948}
}

@inproceedings{pei2023invariants,
  author = {Kexin Pei and David Bieber and Kensen Shi and Charles Sutton and Pengcheng Yin},
  title = {Can Large Language Models Reason about Program Invariants?},
  booktitle = {Proceedings of the 40th International Conference on Machine Learning},
  series = {Proceedings of Machine Learning Research},
  volume = {202},
  pages = {27496--27520},
  year = {2023},
  publisher = {PMLR},
  url = {https://proceedings.mlr.press/v202/pei23a.html}
}

@inproceedings{chakaraborty2023ranking,
  author = {Saikat Chakraborty and Shuvendu K. Lahiri and Sarah Fakhoury and Madanlal Musuvathi and Akash Lal and Aseem Rastogi and Aditya Senthilnathan and Rahul Sharma and Nikhil Swamy},
  title = {Ranking {LLM}-Generated Loop Invariants for Program Verification},
  booktitle = {Findings of the Association for Computational Linguistics: EMNLP 2023},
  month = dec,
  year = {2023},
  address = {Singapore},
  publisher = {Association for Computational Linguistics},
  pages = {9164--9175},
  doi = {10.18653/v1/2023.findings-emnlp.614},
  url = {https://aclanthology.org/2023.findings-emnlp.614/}
}

@inproceedings{wu2024lam4inv,
  author = {Guangyuan Wu and Weining Cao and Yuan Yao and Hengfeng Wei and Taolue Chen and Xiaoxing Ma},
  title = {{LLM} Meets Bounded Model Checking: Neuro-symbolic Loop Invariant Inference},
  booktitle = {Proceedings of the 39th IEEE/ACM International Conference on Automated Software Engineering},
  series = {ASE '24},
  pages = {406--417},
  year = {2024},
  doi = {10.1145/3691620.3695014}
}

@article{wei2026quokka,
  author = {Anjiang Wei and Tianran Sun and Tarun Suresh and Haoze Wu and Ke Wang and Alex Aiken},
  title = {Quokka: Accelerating Program Verification with {LLMs} via Invariant Synthesis},
  journal = {arXiv preprint arXiv:2509.21629},
  year = {2025},
  doi = {10.48550/arXiv.2509.21629}
}

@article{bhat2024synver,
  author = {Prasita Mukherjee and Benjamin Delaware},
  title = {Towards Automated Verification of {LLM}-Synthesized {C} Programs},
  journal = {arXiv preprint arXiv:2410.14835},
  year = {2024},
  doi = {10.48550/arXiv.2410.14835}
}

@inproceedings{lv2026aquaforte,
  author = {Kunhang Lv and Yuhang Dong and Rui Han and Fuqi Jia and Feifei Ma and Jian Zhang},
  title = {{LLM}-Guided Quantified {SMT} Solving over Uninterpreted Functions},
  booktitle = {Proceedings of the AAAI Conference on Artificial Intelligence},
  volume = {40},
  number = {17},
  pages = {14304--14312},
  year = {2026},
  doi = {10.1609/aaai.v40i17.38445},
  url = {https://ojs.aaai.org/index.php/AAAI/article/view/38445}
}

@inproceedings{feng2026llminductive,
  author = {Weizhi Feng and Shidong Shen and Jiaxiang Liu and Taolue Chen and Fu Song and Zhilin Wu},
  title = {Can {LLM} Aid in Solving Constraints with Inductive Definitions?},
  booktitle = {Formal Methods -- 27th International Symposium, {FM} 2026, Tokyo, Japan, May 18--22, 2026, Proceedings, Part {II}},
  editor = {Augusto Sampaio and Mari{\"e}lle Stoelinga},
  series = {Lecture Notes in Computer Science},
  volume = {16557},
  pages = {111--132},
  publisher = {Springer},
  year = {2026},
  doi = {10.1007/978-3-032-26220-2_6},
  url = {https://doi.org/10.1007/978-3-032-26220-2_6}
}

@inproceedings{jiang2023dsp,
  author = {Albert Qiaochu Jiang and Sean Welleck and Jin Peng Zhou and Timoth{\'e}e Lacroix and Jiacheng Liu and Wenda Li and Mateja Jamnik and Guillaume Lample and Yuhuai Wu},
  title = {Draft, Sketch, and Prove: Guiding Formal Theorem Provers with Informal Proofs},
  booktitle = {The Eleventh International Conference on Learning Representations},
  year = {2023},
  url = {https://openreview.net/forum?id=SMa9EAovKMC}
}

@inproceedings{first2023baldur,
  author = {Emily First and Markus N. Rabe and Talia Ringer and Yuriy Brun},
  title = {Baldur: Whole-Proof Generation and Repair with Large Language Models},
  booktitle = {Proceedings of the 31st ACM Joint European Software Engineering Conference and Symposium on the Foundations of Software Engineering},
  series = {ESEC/FSE '23},
  pages = {1229--1241},
  year = {2023},
  doi = {10.1145/3611643.3616243}
}

@inproceedings{yang2023leandojo,
  author = {Kaiyu Yang and Aidan M. Swope and Alex Gu and Rahul Chalamala and Peiyang Song and Shixing Yu and Saad Godil and Ryan J. Prenger and Animashree Anandkumar},
  title = {{LeanDojo}: Theorem Proving with Retrieval-Augmented Language Models},
  booktitle = {Advances in Neural Information Processing Systems},
  volume = {36},
  pages = {21573--21612},
  year = {2023},
  url = {https://papers.nips.cc/paper_files/paper/2023/hash/4441469427094f8873d0fecb0c4e1cee-Abstract-Datasets_and_Benchmarks.html}
}

@inproceedings{thakur2024copra,
  author = {Amitayush Thakur and George Tsoukalas and Yeming Wen and Jimmy Xin and Swarat Chaudhuri},
  title = {An In-Context Learning Agent for Formal Theorem-Proving},
  booktitle = {First Conference on Language Modeling},
  year = {2024},
  url = {https://openreview.net/forum?id=V7HRrxXUhN}
}

@article{yang2025autoverus,
  author = {Chenyuan Yang and Xuheng Li and Md Rakib Hossain Misu and Jianan Yao and Weidong Cui and Yeyun Gong and Chris Hawblitzel and Shuvendu K. Lahiri and Jacob R. Lorch and Shuai Lu and Fan Yang and Ziqiao Zhou and Shan Lu},
  title = {{AutoVerus}: Automated Proof Generation for {Rust} Code},
  journal = {Proceedings of the ACM on Programming Languages},
  volume = {9},
  number = {OOPSLA2},
  pages = {3454--3482},
  year = {2025},
  doi = {10.1145/3763174}
}

@article{tu2026autorocq,
  author = {Haoxin Tu and Huan Zhao and Yahui Song and Mehtab Zafar and Ruijie Meng and Abhik Roychoudhury},
  title = {Agentic Verification of Software Systems},
  journal = {Proceedings of the ACM on Software Engineering},
  volume = {3},
  number = {FSE},
  pages = {3558--3581},
  year = {2026},
  doi = {10.1145/3808164}
}

@article{liu2026kverus,
  author = {Yuwei Liu and Xinyi Wan and Yanhao Wang and Minghua Wang and Lin Huang and Tao Wei},
  title = {{KVerus}: Scalable and Resilient Formal Verification Proof Generation for {Rust} Code},
  journal = {arXiv preprint arXiv:2605.03822},
  year = {2026},
  doi = {10.48550/arXiv.2605.03822}
}

@incollection{floyd1967assigning,
  author = {Robert W. Floyd},
  title = {Assigning Meanings to Programs},
  booktitle = {Mathematical Aspects of Computer Science},
  editor = {J. T. Schwartz},
  series = {Proceedings of Symposia in Applied Mathematics},
  volume = {19},
  pages = {19--32},
  publisher = {American Mathematical Society},
  year = {1967},
  doi = {10.1090/psapm/019/0235771}
}

@article{hoare1969axiomatic,
  author = {C. A. R. Hoare},
  title = {An Axiomatic Basis for Computer Programming},
  journal = {Communications of the ACM},
  volume = {12},
  number = {10},
  pages = {576--580},
  year = {1969},
  doi = {10.1145/363235.363259}
}

@article{dijkstra1975guarded,
  author = {Edsger W. Dijkstra},
  title = {Guarded Commands, Nondeterminacy and Formal Derivation of Programs},
  journal = {Communications of the ACM},
  volume = {18},
  number = {8},
  pages = {453--457},
  year = {1975},
  doi = {10.1145/360933.360975}
}

@article{zhao2026lemmanet,
  author = {Huan Zhao and Haoxin Tu and Zhengyao Liu and Martin Rinard and Abhik Roychoudhury},
  title = {Lemma Discovery in Agentic Program Verification},
  journal = {arXiv preprint arXiv:2603.22114},
  year = {2026},
  doi = {10.48550/arXiv.2603.22114}
}

@inproceedings{beyer2025svcomp,
  author = {Dirk Beyer and Jan Strej{\v{c}}ek},
  title = {Improvements in Software Verification and Witness Validation: {SV-COMP} 2025},
  booktitle = {Tools and Algorithms for the Construction and Analysis of Systems},
  series = {Lecture Notes in Computer Science},
  volume = {15698},
  pages = {151--186},
  publisher = {Springer},
  year = {2025},
  doi = {10.1007/978-3-031-90660-2_9}
}

@inproceedings{beyer2009cpachecker,
  author = {Dirk Beyer and M. Erkan Keremoglu},
  title = {{CPAchecker}: A Tool for Configurable Software Verification},
  booktitle = {Computer Aided Verification},
  series = {Lecture Notes in Computer Science},
  volume = {6806},
  pages = {184--190},
  publisher = {Springer},
  year = {2011},
  doi = {10.1007/978-3-642-22110-1_16}
}

@misc{kroening2023cbmc,
  author = {Daniel Kroening and Peter Schrammel and Michael Tautschnig},
  title = {{CBMC}: The {C} Bounded Model Checker},
  year = {2023},
  eprint = {2302.02384},
  archivePrefix = {arXiv},
  primaryClass = {cs.SE},
  doi = {10.48550/arXiv.2302.02384},
  url = {https://arxiv.org/abs/2302.02384}
}

@inproceedings{cadar2008klee,
  author = {Cristian Cadar and Daniel Dunbar and Dawson Engler},
  title = {{KLEE}: Unassisted and Automatic Generation of High-Coverage Tests for Complex Systems Programs},
  booktitle = {Proceedings of the 8th USENIX Symposium on Operating Systems Design and Implementation},
  series = {OSDI '08},
  pages = {209--224},
  year = {2008}
}

@inproceedings{liu2020vulnerabilitydistribution,
  author = {Bingchang Liu and Guozhu Meng and Wei Zou and Qi Gong and Feng Li and Min Lin and Dandan Sun and Wei Huo and Chao Zhang},
  title = {A Large-Scale Empirical Study on Vulnerability Distribution Within Projects and the Lessons Learned},
  booktitle = {Proceedings of the 42nd International Conference on Software Engineering},
  pages = {1547--1559},
  publisher = {ACM},
  year = {2020},
  doi = {10.1145/3377811.3380923}
}

@inproceedings{demoura2008z3,
  author = {Leonardo de Moura and Nikolaj Bj{\o}rner},
  title = {{Z3}: An Efficient {SMT} Solver},
  booktitle = {Tools and Algorithms for the Construction and Analysis of Systems},
  series = {Lecture Notes in Computer Science},
  volume = {4963},
  pages = {337--340},
  publisher = {Springer},
  year = {2008},
  doi = {10.1007/978-3-540-78800-3_24}
}

@inproceedings{leino2010dafny,
  author = {K. Rustan M. Leino},
  title = {{Dafny}: An Automatic Program Verifier for Functional Correctness},
  booktitle = {Logic for Programming, Artificial Intelligence, and Reasoning},
  series = {Lecture Notes in Computer Science},
  volume = {6355},
  pages = {348--370},
  publisher = {Springer},
  year = {2010},
  doi = {10.1007/978-3-642-17511-4_20}
}

@incollection{hahnle2019deductive,
  author = {Reiner H{\"a}hnle and Marieke Huisman},
  title = {Deductive Software Verification: From Pen-and-Paper Proofs to Industrial Tools},
  booktitle = {Computing and Software Science: State of the Art and Perspectives},
  series = {Lecture Notes in Computer Science},
  volume = {10000},
  pages = {345--373},
  publisher = {Springer},
  year = {2019},
  doi = {10.1007/978-3-319-91908-9_18}
}

@inproceedings{leino2016triggers,
  author = {K. Rustan M. Leino and Cl{\'e}ment Pit-Claudel},
  title = {Trigger Selection Strategies to Stabilize Program Verifiers},
  booktitle = {Computer Aided Verification},
  series = {Lecture Notes in Computer Science},
  volume = {9779},
  pages = {361--381},
  publisher = {Springer},
  year = {2016},
  doi = {10.1007/978-3-319-41528-4_20}
}

@inproceedings{zhou2024contextpruning,
  author = {Yi Zhou and Jay Bosamiya and Jessica G. Li and Marijn J. H. Heule and Bryan Parno},
  title = {Context Pruning for More Robust {SMT}-based Program Verification},
  booktitle = {Formal Methods in Computer-Aided Design},
  pages = {59--69},
  publisher = {TU Wien Academic Press},
  year = {2024},
  doi = {10.34727/2024/isbn.978-3-85448-065-5_12}
}

@misc{li2026agentharness,
  author = {Junjie Li and Xi Xiao and Yunbei Zhang and Chen Liu and Lin Zhao and Xiaoying Liao and Yingrui Ji and Janet Wang and Jianyang Gu and Yingqiang Ge and Weijie Xu and Xi Fang and Xiang Xu and Tianchen Zhao and Youngeun Kim and Tianyang Wang and Jihun Hamm and Smita Krishnaswamy and Jun Huan and Chandan K. Reddy},
  title = {Agent Harness Engineering: A Survey},
  year = {2026},
  howpublished = {\url{https://openreview.net/pdf?id=eONq7FdiHa}}
}

@techreport{barrett2010smtlib,
  author = {Clark Barrett and Aaron Stump and Cesare Tinelli},
  title = {The {SMT-LIB} Standard: Version 2.0},
  institution = {Department of Computer Science, The University of Iowa},
  year = {2010},
  url = {https://smt-lib.org/papers/smt-lib-reference-v2.0-r10.12.21.pdf},
  note = {Release: December 21, 2010}
}

@misc{barrett2025fixedsizebitvectors,
  author = {Clark Barrett and Pascal Fontaine and Silvio Ranise and Cesare Tinelli},
  title = {The {SMT-LIB} Theory of Fixed-Size Bit-Vectors},
  year = {2025},
  howpublished = {\url{https://smt-lib.org/theories-FixedSizeBitVectors.shtml}}
}

@inproceedings{dutertre2006fast,
  author = {Bruno Dutertre and Leonardo de Moura},
  title = {A Fast Linear-Arithmetic Solver for {DPLL(T)}},
  booktitle = {Computer Aided Verification},
  series = {Lecture Notes in Computer Science},
  volume = {4144},
  pages = {81--94},
  publisher = {Springer},
  year = {2006},
  doi = {10.1007/11817963_11}
}

@inproceedings{jovanovic2011cutting,
  author = {Dejan Jovanovi{\'c} and Leonardo de Moura},
  title = {Cutting to the Chase Solving Linear Integer Arithmetic},
  booktitle = {Automated Deduction -- {CADE-23}},
  series = {Lecture Notes in Computer Science},
  volume = {6803},
  pages = {338--353},
  publisher = {Springer},
  year = {2011},
  doi = {10.1007/978-3-642-22438-6_26}
}

@article{griggio2010practical,
  author = {Alberto Griggio},
  title = {A Practical Approach to Satisfiability Modulo Linear Integer Arithmetic},
  journal = {Journal on Satisfiability, Boolean Modeling and Computation},
  volume = {8},
  number = {1--2},
  pages = {1--27},
  year = {2012},
  doi = {10.3233/SAT190086}
}

@inproceedings{chakraborty2017matching,
  author = {Supratik Chakraborty and Ashutosh Gupta and Rahul Jain},
  title = {Matching Multiplications in Bit-Vector Formulas},
  booktitle = {Verification, Model Checking, and Abstract Interpretation},
  series = {Lecture Notes in Computer Science},
  volume = {10145},
  pages = {131--150},
  publisher = {Springer},
  year = {2017},
  doi = {10.1007/978-3-319-52234-0_8}
}

@inproceedings{zohar2022bitprecise,
  author = {Yoni Zohar and Ahmed Irfan and Makai Mann and Aina Niemetz and Andres N{\"o}tzli and Mathias Preiner and Andrew W. Reynolds and Clark Barrett and Cesare Tinelli},
  title = {Bit-Precise Reasoning via Int-Blasting},
  booktitle = {Verification, Model Checking, and Abstract Interpretation},
  series = {Lecture Notes in Computer Science},
  volume = {13182},
  pages = {496--518},
  publisher = {Springer},
  year = {2022},
  doi = {10.1007/978-3-030-94583-1_24}
}

@inproceedings{niemetz2024scalable,
  author = {Aina Niemetz and Mathias Preiner and Yoni Zohar},
  title = {Scalable Bit-Blasting with Abstractions},
  booktitle = {Computer Aided Verification},
  series = {Lecture Notes in Computer Science},
  volume = {14681},
  pages = {178--200},
  publisher = {Springer},
  year = {2024},
  doi = {10.1007/978-3-031-65627-9_9}
}

@article{teuber2020incremental,
  author = {Samuel Teuber and Marko Kleine B{\"u}ning and Carsten Sinz},
  title = {An Incremental Abstraction Scheme for Solving Hard {SMT}-Instances over Bit-Vectors},
  journal = {arXiv preprint arXiv:2008.10061},
  year = {2020},
  doi = {10.48550/arXiv.2008.10061}
}

@inproceedings{jovanovic2017solving,
  author = {Dejan Jovanovi{\'c}},
  title = {Solving Nonlinear Integer Arithmetic with {MCSAT}},
  booktitle = {Verification, Model Checking, and Abstract Interpretation},
  series = {Lecture Notes in Computer Science},
  volume = {10145},
  pages = {330--346},
  publisher = {Springer},
  year = {2017},
  doi = {10.1007/978-3-319-52234-0_18}
}

@article{backeman2021interpolating,
  author = {Peter Backeman and Philipp R{\"u}mmer and Aleksandar Zelji{\'c}},
  title = {Interpolating Bit-Vector Formulas Using Uninterpreted Predicates and {Presburger} Arithmetic},
  journal = {Formal Methods in System Design},
  volume = {57},
  number = {2},
  pages = {121--156},
  year = {2021},
  doi = {10.1007/s10703-021-00372-6}
}

@misc{dafnyref,
  author = {{The dafny-lang community}},
  title = {Dafny Reference Manual},
  year = {2026},
  howpublished = {\url{https://dafny.org/dafny/DafnyRef/DafnyRef}},
  note = {Accessed: 2026-07-01}
}

@article{kirchner2015framac,
  author = {Florent Kirchner and Nikolai Kosmatov and Virgile Prevosto and Julien Signoles and Boris Yakobowski},
  title = {{Frama-C}: A Software Analysis Perspective},
  journal = {Formal Aspects of Computing},
  volume = {27},
  number = {3},
  pages = {573--609},
  year = {2015},
  doi = {10.1007/s00165-014-0326-7}
}

@misc{svcomp2026results,
  author = {{SV-COMP 2026 Organizers}},
  title = {{SV-COMP 2026} Results Verification},
  year = {2026},
  howpublished = {\url{https://sv-comp.sosy-lab.org/2026/results/results-verified/}}
}

@inproceedings{filliatre2013why3,
  author = {Jean-Christophe Filli{\^a}tre and Andrei Paskevich},
  title = {{Why3} --- Where Programs Meet Provers},
  booktitle = {Programming Languages and Systems},
  series = {Lecture Notes in Computer Science},
  volume = {7792},
  pages = {125--128},
  publisher = {Springer},
  year = {2013},
  doi = {10.1007/978-3-642-37036-6_8}
}

@manual{baudin2008acsl,
  author = {Patrick Baudin and Jean-Christophe Filli{\^a}tre and Claude March{\'e} and Benjamin Monate and Yannick Moy and Virgile Prevosto},
  title = {{ACSL}: {ANSI/ISO C} Specification Language. Preliminary Design, version 1.4},
  year = {2008},
  note = {\url{https://www.frama-c.com/html/acsl.html}}
}

\end{document}